\documentclass[reqno]{amsart}

\newcommand{\field}[1]{{\mathbb{#1}}}
\newcommand{\cz}{\field{C}}
\newcommand{\diag}{\operatorname{diag}}

\newcommand{\e}{\hbox{\rm e}}

\newcommand{\bb}[1]{{\mathbb{#1}}}
\newcommand{\bbN}{{\mathbb{N}}}
\newcommand{\bbR}{{\mathbb{R}}}

\newcommand{\bbC}{{\mathbb{C}}}

\newcommand{\calK}{{\mathcal K}}
\newcommand{\calR}{{\mathcal R}}
\newcommand{\no}{\nonumber}
\newcommand{\lb}{\label}
\newcommand{\ul}{\underline}

\newcommand{\kdv}{\operatorname{KdV}}

\newcommand{\g}{g}
\renewcommand{\Re}{\text{\rm Re}}

\DeclareMathOperator{\sKdV}{s-KdV}
\DeclareMathOperator{\KP}{KP}

\allowdisplaybreaks

\numberwithin{equation}{section}

\newtheorem{theorem}{Theorem}[section]
\newtheorem{lemm}[theorem]{Lemma}
\newtheorem{defi}[theorem]{Definition}
\newtheorem{rem}[theorem]{Remark}
\newtheorem{exa}[theorem]{Example}

\begin{document}

\title{On a Theorem of Halphen and its Application \newline to Integrable
Systems}

   \thanks{Based upon work supported by the
US National Science Foundation under Grant No.~DMS-9970299. \\
\it{J. Math. Anal. Appl. {\bf 251}, 504--526 (2000).}}

\author{F.~Gesztesy}
\address{Department of Mathematics,
University of Missouri, Columbia, MO 65211, USA}
\email{fritz@math.missouri.edu}
\urladdr{http://www.math.missouri.edu/people/fgesztesy.html}
\author{K.~Unterkofler}
\address{Department of Computer Science,
Applied Mathematics Group,
FH-Vorarlberg,
A--6850 Dornbirn,
Austria}
\email{karl.unterkofler@fh-vorarlberg.ac.at}
\author{R.~Weikard}
\address{Department of Mathematics,
University of Alabama at Birmingham, \\
Birmingham, AL  35294--1170, USA}
\email{rudi@math.uab.edu}
\urladdr{http://www.math.uab.edu/rudi}


\begin{abstract}
We extend Halphen's theorem which characterizes the solutions
of certain $n$th-order differential equations with rational
coefficients and meromorphic fundamental systems to a
first-order $n \times n$ system of differential equations.
As an application of this circle of ideas we consider stationary
rational algebro-geometric solutions of the $\kdv$ hierarchy and
illustrate some of the connections with completely integrable
models of the Calogero-Moser-type. In particular, our treatment recovers
the  complete characterization of the isospectral class of such rational
KdV solutions in terms of a precise description of the
Airault-McKean-Moser locus of their poles.
\end{abstract}


\maketitle

\section{Introduction} \lb{s1}
The purpose of this paper is twofold. First we prove an extension of
Halphen's theorem, which characterizes the fundamental system of
solutions of certain $n$th-order ordinary differential equations with
rational coefficients to first-order $n\times n$ systems. In the second
part of this paper we show how to apply Halphen's theorem to completely
integrable systems of the Calogero-Moser-type, recovering a
complete  characterization of the isospectral class of all
algebro-geometric rational solutions of the KdV hierarchy.

We start by describing Halphen's original result. Consider the following
$n$th-order differential equation
\begin{align}
   q_n(z)y^{(n)}(z)+q_{n-1}(z) y^{(n-1)}(z)+\dots+
q_0(z) y(z) =0, \label{1.18}
   \end{align}
where $q_j(z)$ are polynomials, and the order of $q_n(z)$ is at
least the order of $q_j(z)$ for  all  $0\leq j\leq (n-1)$, that is,
\begin{subequations} \lb{1.20}
\begin{align}
& q_m(z) \text{ are polynomials, $0\leq m\leq n$,} \lb{1.20a} \\
& q_m(z)/q_n(z) \text{ are bounded near $\infty$ for all
$0\leq m \leq n-1$}.
\lb{1.20b}
\end{align}
\end{subequations}
Then the  zeros of $q_n(z)$ are the possible singularities of solutions
of \eqref{1.18}.

Assuming the fundamental system of solutions of \eqref{1.18} to be
meromorphic, the following theorem due to Halphen holds.
\begin{theorem} {\rm (}Halphen \cite{Ha85}, Ince
\cite[p.~372--375]{In56}{\rm)} \lb{t1.6}
Assume \eqref{1.20} and
suppose \eqref{1.18} has a meromorphic fundamental system of
solutions. Then the general solution of \eqref{1.18} is of the form
\begin{align}
y (z ) =  \sum_{m=1}^{n} c_m r_m(z) e^{\lambda_m z}, \lb{1.21}
   \end{align}
where   $r_m(z)$ are rational functions of $z$,
$\lambda_m\in\bbC$, $1\leq m\leq n$, and $c_m$, $1\leq m\leq n$
are arbitrary complex constants.
\end{theorem}
Moreover, the converse of Halphen's theorem holds as well.
\begin{theorem} {\rm (}Ince \cite[p.~374--375]{In56}{\rm)} \lb{t1.6a}
Suppose $r_m(z)$ are rational functions of $z$ and
$\lambda_m, c_m\in\bbC$, $1\leq m\leq n$. If $r_1(z)e^{\lambda_1 z},
\dots,r_n(z)e^{\lambda_n z}$ are linearly independent, then
\begin{align}
y (z ) =  \sum_{m=1}^{n} c_m r_m(z) e^{\lambda_m z} \lb{1.22}
\end{align}
is the general solution of an $n$th-order equation of the type
\eqref{1.18}, whose  coefficients satisfy \eqref{1.20}.
\end{theorem}
\begin{rem} \lb{r1.7}
{\em We note that Halphen's main idea of proof in
\cite{Ha85} consists of replacing the rational coefficients in
\eqref{1.18} by appropriate elliptic coefficients {\rm(}as discussed in
\cite{Ha84}{\rm)} followed by an application of Picard's theorem
{\rm(}cf., e.g., \cite[p.~375--378]{In56}{\rm)}. A closer examination of
his argument seems to reveal a lack of proof of the crucial fact that the
associated differential equation with elliptic coefficients necessarily
has a meromorphic fundamental system of solutions. A proof of
Theorem~\ref{t1.6} {\rm(}and Theorem~\ref{t1.6a}{\rm)}, using  a
completely different strategy, is provided in Ince's monograph
\cite[p.~372--375]{In56}. }
\end{rem}
One of the principal aims of this note is to prove a first-order
$n\times n$ system generalization of Halphen's Theorem~\ref{t1.6}
and its converse, Theorem~\ref{t1.6a}, in Section~\ref{s2}.

Analogous results hold for $n$th-order equations and first-order systems
with periodic and elliptic coefficients. For a glimpse at the vast
literature in these cases and their applications to completely integrable
systems we refer the interested reader to
\cite{GS98}--\cite{GW98a}, \cite{We99}, \cite{We00} and the literature
therein.

In Section~\ref{s3} we then apply Halphen's theorem to the problem of
characterizing the isospectral class of all stationary rational KdV
solutions. All such (nonconstant) solutions $q$ are well-known to be
necessarily of the form
\begin{equation}
q(z)=q_\infty-\sum_{\ell=1}^M s_\ell(s_\ell+1)(z-\zeta_\ell)^{-2}
\lb{1.23}
\end{equation}
for some $q_\infty\in\bbC$, $\{\zeta_\ell\}_{1\leq\ell\leq
M}\subset\bbC$, $\zeta_\ell^\prime\neq\zeta_\ell$ for
$\ell^\prime\neq\ell$, and
\begin{equation}
s_\ell\in\bbN, \,\,1\leq\ell\leq M
\text{ with } \sum_{\ell=1}^M s_\ell(s_\ell+1)=g(g+1) \lb{1.24}
\end{equation}
for some $g\in\bbN$, and the underlying spectral curve is then of the
especially simple rational type
\begin{equation}
y^2=(E-q_\infty)^{2g+1}. \lb{1.25}
\end{equation}
On the other hand, not every
$q$ of the type  \eqref{1.23}, \eqref{1.24} is an algebro-geometric
solution of the KdV hierarchy. In general, the points $\zeta_\ell$ must
satisfy a set of intricate constraints. In fact, necessary and
sufficient  conditions on $\zeta_\ell$ for $q$ in \eqref{1.23} to be a
rational KdV solution are given by
\begin{equation}
\sum_{\substack{\ell^\prime=1\\ \ell^\prime\neq \ell}}^M
\frac{s_{\ell^\prime}(s_{\ell^\prime}+1)}{(\zeta_{\ell}
-\zeta_{\ell^\prime})^{2k+1}}=0 \quad
   \text{for $k=1, ..., s_{\ell^\prime}$ and $\ell=1,\dots,M$.} \lb{1.26}
\end{equation}
This result was first derived by Duistermaat and Gr\"unbaum
\cite{DG86} (cf.~p.~199) in 1986, as a by-product of their investigations
of bispectral pairs of differential operators. We will provide an
elementary  derivation of this result on the basis of Halphen's theorem
and an explicit Frobenius-type analysis in Section~\ref{s3}.

For a fixed $g\in\bbN$, \eqref{1.24} and
\eqref{1.26} yield a complete parametrization of all rational KdV
solutions belonging to the  spectral curve \eqref{1.25}. In other words,
they provide a complete  characterization of the isospectral class of KdV
solutions corresponding to \eqref{1.25}. The constraints \eqref{1.26}
represent the proper  generalization of the locus of poles introduced by
Airault, McKean, and Moser \cite{AMM77} in the sense that they explicitly
describe the  situation where poles are permitted to collide (i.e., where
some of  the $s_\ell >1$).

\section{Halphen's theorem for first-order systems} \lb{s2}
This section is devoted to a generalization of Halphen's theorem (and its
converse) to first-order systems.

We briefly describe some of the notation used in this section.
   $I_n$ denotes the identity in
$\bbC^n$. An $m\times m$ diagonal matrix $D=(d_j\delta_{j,k}
)_{1\leq j,k\leq m}$ will occasionally be denoted by
$\diag(d_1,\dots,d_m)$. The operation of transposition is denoted
by the superscript $t$. Moreover, it will be convenient to denote
the set of all
$m\times n$ matrices whose entries are rational functions with respect
to $z\in\bbC$ by $\calR^{m\times n}$, the subset of
$\calR^{m\times n}$ with rational entries bounded at infinity by
$\calR^{m\times n}_\infty$.

We recall that for $T\in\calR^{n\times n}$ invertible and
differentiable with
respect to $z$, the transformation $y(z)=T(z)u(z)$ turns the
first-order system of differential equations $y'(z)=A(z)y(z)$ into the
system $u'(z)=B(z)u(z)$, with $B(z)=T(z)^{-1}(A(z)T(z)-T'(z))$.
\begin{defi} \lb{d2.1}
(i) Two matrices $A, B\in\calR^{n\times n}$ are called \textit{of the
same kind} if
there exists an invertible matrix $T\in\calR^{n\times n}$  such that
\begin{align}
B(z)=T(z)^{-1}(A(z)T(z)-T'(z)). \label{}
\end{align}
(ii) $B\in\calR^{n\times n}$ is called {\it reduced of order $k$} if
$B_{j,\ell}=\delta_{j+1,\ell}$ for all $1\leq j\leq k$ and $1\leq
\ell\leq n$.
\end{defi}
Our approach, including the notion of matrices being ``of the same
kind'', was inspired by Loewy \cite{Lo18}. The relation of being of the
same kind is obviously an equivalence relation on $\calR^{n\times n}$.
The relation of being of the same kind is obviously an equivalence
relation on $\calR^{n\times n}$.
\begin{lemm} \label{l2.3}
Suppose that $A\in\calR^{n\times n}_{\infty}$ is reduced of order
$k-1$. Then either
$A_{k,k+1} = \dots\ =A_{k,n}=0$, or else there exists a matrix
$B\in\calR^{n\times n}_{\infty}$ of the same kind as $A$ and also
reduced of order $k-1$ but with the additional property that
$B_{k,k+1}(\infty)\neq 0$. Moreover,
$A(\infty)$ and $B(\infty)$ have the same
eigenvalues counting algebraic multiplicities.
\end{lemm}

\begin{proof} We assume not all of the entries $A_{k,k+1}$,\dots,
$A_{k,n}$ are equal to zero. Consider the $(n-k+1)\times(n-k)$ matrix
in the lower right corner of $A$ and denote it by $R$. Suppose that
$r$ is the largest nonnegative integer such that $z^r R_{1,j}(z)$ remains
bounded near infinity for every $j\in\{1,\dots,n-k\}$. Then there exists
an
$\ell\in\{1,\dots,n-k\}$ such that $z^r R_{1,\ell}(z)$ does not vanish
at infinity. Denote the constant $(n-k)\times(n-k)$ matrix, which
achieves the exchange of columns $1$ and $\ell$ of $R(z)$, by $C$. Then
the first row of $z^r R(z)C$ is bounded at infinity and the first entry
in that row does not vanish at infinity. Next, define
\begin{align}
T(z)=\begin{pmatrix}I_k&0\\0&z^rC\end{pmatrix},
\label{}
\end{align}
where $I_k$ is the $k\times k$ identity matrix. Let
\begin{align}
A(z)=\begin{pmatrix}
\tilde A_{1,1}(z)&\tilde A_{1,2}(z)\\ \tilde A_{2,1}(z)&\tilde
A_{2,2}(z) \end{pmatrix}, \label{}
\end{align}
where $\tilde A_{1,1}(z)$ and $\tilde A_{2,2}(z)$ are square matrices
with $k$ and $n-k$ rows, respectively. Then
\begin{align}
T(z)^{-1}A(z)T(z)=\begin{pmatrix}
\tilde A_{1,1}(z)&z^r \tilde A_{1,2}(z)C\\ z^{-r} C^{-1} \tilde
A_{2,1}(z) &C^{-1} \tilde A_{2,2}(z) C \end{pmatrix}. \label{}
\end{align}
Since only the last row of $\tilde A_{1,2}(z)$ is different from zero,
and since that row equals the first row of $R(z)$, the matrix
$T(z)^{-1}A(z)T(z)$ remains bounded at infinity and its first $k-1$
rows are the same as those of $A(z)$. The matrix $C$ was chosen so that
the first entry in the last row of $z^r\tilde A_{1,2}(z)C$ does not
vanish at infinity. Since
\begin{equation}
\lim_{z\to\infty}T(z)^{-1}T'(z)=0, \lb{2.5}
\end{equation}
we conclude that $B=T^{-1}(AT-T')\in\calR^{n\times n}_{\infty}$ is
reduced of order $k-1$ and that $B_{k,k+1}(\infty)\neq0$.

Finally we prove that $A(\infty)$ and $B(\infty)$ have the same
eigenvalues counting algebraic multiplicities. Since $T(\infty)$ might not
exist, we first compute
\begin{align}
   &\det(\lim_{z\to\infty}((T^{-1}AT)(z)-\lambda I_n))
   =\lim_{z\to\infty}\det((T^{-1}AT)(z)-\lambda I_n) \nonumber\\
   &=\lim_{z\to\infty}\det(A(z)-\lambda I_n)
   =\det(\lim_{z\to\infty}(A(z)-\lambda I_n)). \lb{2.6}
\end{align}
By \eqref{2.5}, the left-hand side of \eqref{2.6} is the characteristic
polynomial of $B(\infty)$, while the right-hand side is the
characteristic polynomial of $A(\infty)$. This completes the proof.
\end{proof}
\begin{lemm} \label{l2.4}
Assume that $A\in\calR^{n\times n}_{\infty}$ is reduced of order $k-1$ and
suppose that
   $A_{k,k+1}(\infty)\neq 0$. Then
there exists a matrix $B\in\calR^{n\times n}_{\infty}$ of the same
kind as $A$, which is reduced of order $k$.  Moreover, $A(\infty)$ and
$B(\infty)$ are similar and hence isospectral {\rm(}i.e., their
eigenvalues, including algebraic and geometric multiplicities,
coincide{\rm)}.
\end{lemm}
\begin{proof}
Let $T\in\calR^{n\times n}$ denote the $n\times n$ matrix obtained from
the identity matrix $I_n$ by replacing its $(k+1)$st row by
\begin{align}
(-A_{k,1},\dots,-A_{k,k},1,
   -A_{k,k+2},\dots,-A_{k,n})/A_{k,k+1}. \label{}
\end{align}
$T^{-1}$ is then the matrix obtained from the identity
matrix $I_n$ by replacing the $(k+1)$st row by
$(A_{k,1},\dots,A_{k,n})$.
Note that the entries of $T$ and $T^{-1}$ are rational and
bounded at infinity. Hence the matrix
$B=T^{-1}(AT-T')$ has rational entries
bounded at infinity. A straightforward calculation then shows that the
first $k$ rows of $B$ have the desired form. Since $T$ and
$T'$ are bounded at infinity, $\lim_{z\to\infty}T(z)^{-1}T'(z)=0$
and hence $B(\infty) =T(\infty)^{-1}A(\infty)T(\infty)$.
\end{proof}
\begin{theorem} \label{t2.5}
Let $Q\in\calR_\infty^{n\times n}$ and suppose that the
first-order system
$y'(z)=Q(z)y(z)$ has a meromorphic fundamental system of solutions.
Then $y'=Qy$ has a fundamental matrix of the type
\begin{align}
Y(z)=R(z)\exp(\diag(\lambda_1 z,\dots,\lambda_n z)), \label{}
\end{align}
where $\lambda_1,\dots,\lambda_n$ are the eigenvalues of
$Q(\infty)$ and $R\in\calR^{n\times n}$.
\end{theorem}
\begin{proof}
The theorem will be proved by induction on $n$. Let $n=1$. Any
pole of $Q(z)$ must be of first-order with an integer
residue, that is,
\begin{align}
Q(z)=\lambda_1+\sum_{\ell=1}^N \frac{m_\ell}{z-a_\ell}, \label{}
\end{align}
with $m_1,\dots,m_N\in\bb Z$. Then
$Y(z)=\prod_{\ell=1}^N (z-a_\ell)^{m_\ell} \exp(\lambda_1 z)$
proves the claim for $n=1$.

Next, let $n$ be any natural number and assume that Theorem~\ref{t2.5}
has been proven for any natural number strictly less than $n$.\\
By hypothesis, $Q\in\calR^{n\times n}_{\infty}$ and $Q(z)$ can be
regarded to be reduced at least of order zero. We denote the
eigenvalues of $Q(\infty)$ by $\lambda_1,\dots,\lambda_n$.
Repeated, perhaps alternating, applications of Lemmas~\ref{l2.3} and
\ref{l2.4} then yield the existence of an integer
$k\in\{1,\dots,n\}$, a $k\times k$ matrix
$B_1(z)$, an $(n-k)\times k$ matrix $B_3(z)$, and an $(n-k)\times(n-k)$
matrix $B_4(z)$, such that
\begin{align}
B(z)=\begin{pmatrix}B_1(z)&0\\ B_3(z)&B_4(z)\end{pmatrix}
\label{}
\end{align}
has the following properties:
\begin{enumerate}
\item $B\in\calR_\infty^{n\times n}$.
\item $B$ is of the same kind as $Q$, that is, there exists an
invertible matrix $T\in\calR^{n\times n}$ such that
$B(z)=T(z)^{-1}(Q(z)T(z)-T'(z))$.
\item $B_1(z)$ is reduced of order $k-1$.
\item After a suitable relabeling of the eigenvalues of $Q(\infty)$ the
eigenvalues of $B_1(\infty)$ are $\lambda_1,\dots,\lambda_k$ and the
eigenvalues of $B_4(\infty)$ are $\lambda_{k+1},\dots,\lambda_n$.
\item The first-order system $u'(z)=B(z) u(z)$ has a meromorphic
fundamental system of solutions with respect to $z\in\bbC$.
\end{enumerate}
We now have to distinguish whether $k=n$ or $k<n$. In the case $k=n$,
$B(z)=B_1(z)$, and the system $u'(z)=B(z)u(z)$ is equivalent to the
scalar equation
\begin{align}
u_1^{(n)}(z)=B_{n,1}(z)u_1+\dots+B_{n,n}(z) u_1^{(n-1)}(z).
\label{}
\end{align}
In this case Halphen's theorem, Theorem~\ref{t1.6}, and the relations
$u_k(z)=u_1^{(k-1)}(z)$ and $y(z)=T(z)u(z)$ prove our claim.\\
Next, assume that $k<n$. If $w$ is any solution of $w'(z)=B_1(z)w(z)$,
choose a solution $v$ of the nonhomogeneous system
\begin{equation}
v'(z)-B_4(z) v(z)=B_3(z)w(z). \lb{2.14}
\end{equation}
Then
$u=(w,v)^t$ is a solution of $u'(z)=B(z) u(z)$ and hence meromorphic.
Thus every solution of $w'(z)=B_1(z) w(z)$ and, choosing $w(z)=0$
in \eqref{2.14}, also every solution of
$v'(z)=B_4(z)v(z)$ is meromorphic. By the induction hypothesis, there
exist  matrices $R_1\in\calR^{k\times k}$ and $R_4\in\calR^{(n-k)\times
(n-k)}$ such that
\begin{align}
U_1(z)=R_1(z) \diag(\exp(\lambda_1 z),\dots,\exp(\lambda_k z))
\label{}
\end{align}
is a fundamental matrix of $w'(z)=B_1(z) w(z)$ and
\begin{align}
U_4(z)=R_4(z) \diag(\exp(\lambda_{k+1} z),\dots,\exp(\lambda_n z))
\label{}
\end{align}
is a fundamental matrix of $v'(z)=B_4(z) v(z)$.\\
Next define
\begin{align}
U_3(z)=U_4(z)\int^z d\zeta\,U_4^{-1}(\zeta)B_3(\zeta)U_1(\zeta).
\label{}
\end{align}
Then each column of
\begin{align}
U(z)=\begin{pmatrix}U_1(z)&0\\ U_3(z)&U_4(z)\end{pmatrix}
\label{}
\end{align}
is a solution of $u'(z)=B(z)u(z)$ and $U$ is indeed a fundamental
matrix of
$u'(z)=B(z)u(z)$ since $\det(U(z))=\det(U_1(z))\det(U_4(z))\neq0$. It
remains to show that
\begin{align}
U_3(z)=R_3(z) \diag(\exp(\lambda_{1} z),\dots,\exp(\lambda_k z))
\label{}
\end{align}
for some matrix $R_3\in\calR^{(n-k)\times k}$. The entry
in row
$j$ and column $\ell$ of the
matrix $U_4^{-1}(\zeta)B_3(\zeta)U_1(\zeta)$ equals
$\rho_{j,\ell}(\zeta)\exp[(\lambda_\ell-\lambda_{k+j})\zeta]$, where
$\rho_{j,\ell}$ is a rational function, that is,
\begin{align}
\rho_{j,\ell}(\zeta)=\sum_{r=0}^N a_{j,\ell,r} \zeta^r+\sum_{r=1}^M
\sum_{s=1}^{M_r} \frac{b_{j,\ell,r,s}}{(\zeta-z_r)^s}   \label{}
\end{align}
for appropriate choices of the parameters $a_{j,\ell,r}$,
$b_{j,\ell,r,s}$, and pairwise distinct $z_r$. Next we recall that
\begin{equation}
\int^z d\zeta \zeta^s \e^{\lambda \zeta} =
   f(s,\lambda,z) \,\,(\text{mod}{\,(\e^{\lambda z}\bb C(z))}),
\end{equation}
where
\begin{equation}
f(s,\lambda,z)=\begin{cases}
   0 &\text{if $s\geq0$}\\
   \frac{\lambda^{-s-1}}{(-s-1)!}\operatorname{Ei}(\lambda z)
       &\text{if $\lambda\neq0$ and $s\leq-1$}\\
   \ln(z) &\text{if $\lambda=0$ and $s=-1$}\\
   0 &\text{if $\lambda=0$ and $s\leq-2$}
\end{cases}
\end{equation}
and that the exponential integral $\operatorname{Ei(\cdot)}$ has a
logarithmic branch point at zero. Therefore, if
$\lambda_\ell\neq\lambda_{k+j}$,
\begin{align}
   &(U_4^{-1}U_3)_{j,\ell}(z)
=\int^z d\zeta \,\rho_{j,\ell}(\zeta)\e^{(\lambda_\ell
-\lambda_{k+j})\zeta} \label{2.23} \\
   &=\e^{(\lambda_\ell-\lambda_{k+j})z} S_{j,\ell}(z)
   +\sum_{r=1}^M c_{j,\ell,r}
   \e^{(\lambda_\ell-\lambda_{k+j})z_r}
   \operatorname{Ei}((\lambda_\ell-\lambda_{k+j})(z-z_r)), \no
\end{align}
for appropriate rational functions $S_{j,\ell}$ and appropriate complex
numbers $c_{j,\ell,r}$. However, since the entries of $U_3(z)$ and
$U_4(z)$ must be meromorphic, all of the numbers $c_{j,\ell,r}$ must
necessarily vanish. If $\lambda_\ell =\lambda_{k+j}$ a similar
conclusion shows that no logarithmic terms appear so that in either
case $(U_4^{-1}U_3)_{j,\ell}(z)
   \in\e^{(\lambda_\ell-\lambda_{k+j})z}\bb C(z)$.
Hence we obtain
\begin{align}
U_4^{-1}(z)U_3(z)=\diag(\e^{-\lambda_{k+1}z},\dots,\e^{-\lambda_{n}z})
S(z) \diag(\e^{\lambda_{1}z},\dots,\e^{\lambda_{k}z}), \label{}
\end{align}
where $S\in\mathcal R^{(n-k)\times k}$ is the matrix with entries
$S_{j,\ell}$. Thus, $R_3=R_4S\in\mathcal R^{(n-k)\times k}$.
\end{proof}
\begin{rem} \lb{r1.9}
{\em If $Q(\infty) = 0$, the transformation
$z=\frac{1}{\zeta}$
immediately shows that $\zeta=0$, is a regular singular point of
our differential equation. This implies that the fundamental
system at $\zeta=0$ is of the form {\rm(}cf., e.g.,
\cite[Sect.~23]{Wa98}{\rm)}
$Y(\zeta) = U(\zeta) \zeta^m$ for $|\zeta|< \zeta_0$,
where $U(\zeta)$ is a holomorphic matrix for $ |\zeta|< \zeta_0$.
Hence our fundamental system is meromorphic on the whole Riemann sphere
and must  therefore be a purely  rational matrix. }
\end{rem}
Next we consider two examples.
\begin{exa} \lb{e1.11}
The first-order $2\times 2$ system
$$
y'(z) = \left(\begin{matrix}1 & 0\\ z
&  1 +\frac{1}{z} \end{matrix} \right) y (z)
$$
is solved by
$$
   Y(z) =  \left(\begin{matrix} 1   & 1\\
z^2 & z^2 +z \end{matrix} \right)e^z.
$$
\end{exa}
This seems to suggest consideration of even more general systems of the
type $z^{-q} Y'(z) = A(z) Y(z)$, with $q>0$, rather than the case $q=0$
only. But Theorem~\ref{t2.5} can not hold in  general for $q>0$ as
shown by the following elementary counterexample.
\begin{exa} \lb{e1.12}
The first-order $2\times 2$ system
$$
y'(z) =  \left(\begin{matrix}0  & 1\\z^m & 0
\end{matrix} \right) y(z), \quad m\in {\mathbb{N}}
$$
has no solution in terms of elementary functions, although it clearly
has a meromorphic fundamental system. The particular case $m=1$
represents the well-known Airy equation.
\end{exa}
\begin{rem} \lb{r2.6a}
{\em In the case where all eigenvalues $1\leq\lambda_j\leq n$
of $Q(\infty)$ are distinct, we now sketch an alternative proof of
Theorem~\ref{t2.5}, based
on Theorem~12.3 in Wasow's monograph \cite{Wa87}. Since Theorem~12.3 in
\cite{Wa87} only applies to appropriate sectors of the complex plane with
vertex at the origin, we argue as follows. First one can
find a sufficiently small sector $S_3$, which does not contain
any separation
rays. (We recall that a ray (i.e., a half line),  where $\Re(\lambda_j z
-\lambda_k z)=0$ for some pair of distinct integers $j,k$, is called a
separation ray.) Then one chooses two other sectors $S_1,S_2$
with opening
angles $\phi_j < \pi$, $j=1,2$, such that $ S_1 \cup S_2 \cup S_3 =
    \cz\backslash\{0\}$. It is then possible to show that the
transition matrix
from sector $S_1 $ to sector $S_2 $ equals the identity matrix.
Hence, the
solution of the form $  Y (z) =  R(z) \exp({\diag(\lambda_1,
\ldots, \lambda_n)
z})$ in  sector $S_1 $  is valid in sector $S_2$ too and thus can
be continued
into $S_3$ since by hypothesis, the sector $S_3$ contains no
separation rays. }
\end{rem}
Finally, we turn to a converse of Theorem~\ref{t2.5}.
\begin{theorem} \lb{t2.6}
Suppose $R\in\calR^{n\times n}$, $\det(R)\neq 0$, and
$\lambda_1,\dots,\lambda_n\in\bbC$. Then
\begin{align}
Y(z)=R(z)\exp(\diag(\lambda_1 z,\dots,\lambda_n z)) \label{}
\end{align}
is a fundamental matrix of a first-order linear system of differential
equations $y'(z)=Q(z)y(z)$, where $Q\in\calR^{n\times n}$ and
$Q(z)$ is of the
same kind as a matrix in $\calR^{n\times n}_\infty$. In fact, $Q(z)$ is
of the same  kind as the constant diagonal matrix
$\diag(\lambda_1,\dots,\lambda_n)$.
\end{theorem}
\begin{proof} Since
\begin{equation}
Q(z)=R(z)\diag(\lambda_1,\dots,\lambda_n)R(z)^{-1}+R'(z)R(z)^{-1},
\lb{2.27}
\end{equation}
we choose $T(z)=R(z)^{-1}$ and hence obtain $T'=-R^{-1}R'R^{-1}$ and
thus,
\begin{equation}
Q=T^{-1}(\diag(\lambda_1 ,...,\lambda_n)T-T'). \lb{2.28}
\end{equation}
Hence, $Q(z)$ is of the same kind as the constant matrix
$\diag(\lambda_1,\dots,\lambda_n)$.
\end{proof}

\section{Some applications to rational solutions of the stationary
$\kdv$ hierarchy} \lb{s3}

In this section we describe the connections between the preceding
results and infinite-dimensional completely integrable Hamiltonian
systems. For reasons of brevity we will only consider the simplest case
of the $\kdv$ hierarchy, and in accordance with Sections~\ref{s1},
\ref{s2}, only study its stationary rational
solutions bounded at infinity (cf.~\cite{AS78}, \cite{AM78}--\cite{CC77},
\cite{Gr82}, \cite{Ka95}, \cite{Kr78}--\cite{Kr74}, \cite{Mo77},
\cite{Oh88}, \cite{Pe94}, \cite{Sh94}, \cite{So78}, \cite{Wi98} and the
literature cited therein). The principal results on the stationary
$\kdv$ hierarchy  as needed in this section are summarized in the
appendix, and we freely use these results and the notation
established there in what follows.

The rational $\kdv$ solutions bounded at infinity are usually discussed in
a time-dependent setting and the dynamics of their poles is in an
intimate relationship with completely integrable systems of the
Calogero-Moser-type. In our discussion below, the time-dependence will
generally be suppressed and only occasionally be mentioned in connection
with particular isospectral deformations of rational solutions of the
$\kdv$ hierarchy. Our principal focus will be on stationary
(isospectral) aspects of these rational $\kdv$ solutions and the
implications of Halphen's theorem in this context.

We start by quoting a number of known results on stationary rational
$\kdv$ solutions bounded at infinity.
\begin{theorem} \lb{t3.1}
Let $N\in\bbN$ and $\{z_j\}_{1\leq j\leq N}\subset\bbC$. \\
\noindent {\rm (}i{\rm )} {\rm (}Airault, McKean, and Moser
\cite{AMM77}{\rm )}
Any rational solution $q$ of
{\rm(}some, and hence infinitely many equations of\,{\rm)} the $\kdv$
hierarchy, or equivalently, any rational algebro-geometric potential
$q$, is necessarily of the form
\begin{equation}
q(z)=q_\infty -2\sum_{j=1}^N (z-z_j)^{-2},  \lb{3.2}
\end{equation}
for some $q_\infty\in\bbC$ and with $N\in\bbN$ of the special type
$N=g(g+1)/2$ for some
$g\in\bbN$. \\
\noindent {\rm (}ii{\rm )} {\rm (}Airault, McKean, and Moser
\cite{AMM77} {\rm (}see also \cite{We99}{\rm ))} If one allows for
``collisions'' between the
$z_j$, that is, if the set $\{z_j\}_{1\leq j\leq N}$ clusters into
groups of points, then the corresponding rational algebro-geometric
potential $q$ is necessarily of the form
\begin{equation}
q(z)=q_\infty -\sum_{\ell=1}^M s_\ell(s_\ell +1)(z-\zeta_\ell)^{-2},
\lb{3.5}
\end{equation}
where for some $g\in\bbN$,
\begin{subequations} \lb{3.4}
\begin{align}
&\{z_j\}_{1\leq j\leq
N}=\{\zeta_\ell\}_{1\leq\ell\leq M}\subset\bbC, \text{ with $\zeta_\ell$
pairwise distinct,} \lb{3.4a} \\
&  s_\ell\in\bbN,\,\,\, 1\leq\ell \leq M, \no \\
&\sum_{\ell=1}^M s_\ell(s_\ell +1)=2N \text{ for some $N\in\bbN$ of
the type $N=g(g+1)/2$.} \lb{3.4b}
\end{align}
\end{subequations}
\noindent {\rm (}iii{\rm )} The extreme case of all
$z_j$ colliding into one point,  say $\zeta_1$, that is,
$\{z_j\}_{1\leq j\leq N}=\{\zeta_1\}\subset\bbC$
yields an algebro-geometric $\kdv$ potential of the elementary form
\begin{equation}
q(z)=q_\infty -g(g+1)(z-\zeta_1)^{-2}, \quad g\in\bbN \lb{3.6}
\end{equation}
and no additional constraints on $\zeta_1\in\bbC$. \\
\noindent {\rm (}iv{\rm )} In all cases {\rm (}i{\rm )}--{\rm (}iii{\rm
)}, if
$q$ is a rational $\kdv$ potential {\rm (}i.e., if $g\in\bbN$ and the
points $z_j$ {\rm (}resp.
$\zeta_\ell${\rm )} satisfy appropriate restrictions,
cf.~Theorem~\ref{t3.6}{\rm)}, the underlying rational hyperelliptic curve
$\calK_g$ is of the especially simple form
\begin{equation}
\calK_g \colon y^2=(E-q_\infty)^{2g+1}. \lb{3.8}
\end{equation}
In particular, the potentials \eqref{3.2}, \eqref{3.5}, and \eqref{3.6}
are all isospectral {\rm (}assuming \eqref{3.2} and \eqref{3.5} are
algebro-geometric $\kdv$ potentials, of course{\rm )}. \\
\noindent {\rm (}v{\rm )} {\rm (}Weikard \cite{We99}{\rm )} $q$ is a
rational $\kdv$ potential if and only if
$\psi''+(q-E)\psi=0$ has a meromorphic fundamental solutions
{\rm (}w.r.t.~$z${\rm )} for all values of the spectral parameter
$E\in\bbC$.\\
\noindent {\rm (}vi{\rm )} If $q$ is a rational KdV potential of the form
\eqref{3.5}, then $y''+qy=Ey$ has linearly independent solutions of
the Baker-Akhiezer-type
\begin{align}
&\psi_\pm(E,z)=\big(\pm E^{1/2}\big)^{-g}\Bigg(\prod_{j=1}^g
\big(\pm E^{1/2}-\nu_j(z)\big)\Bigg) e^{\pm E^{1/2} z}, \lb{3.8a} \\
&\hspace*{5.53cm} E\in\bbC\backslash\{q_\infty\}, \,\, z\in\bbC, \no
\end{align}
with $\mu_j(z)=\nu_j(z)^2$, $1\leq j\leq g$ the zeros of $F_g(z,x)$
as defined in \eqref{A.10}.
\end{theorem}
\noindent (To avoid annoying case distinctions we will in almost all
circumstances exclude the trivial case $N=g=0$ in this section.)
\begin{rem} \lb{r3.2}
{\em {\rm(}i{\rm)} It must be emphasized that for $N>1$, not any potential
$q$ of the type \eqref{3.2} is
an algebro-geometric $\kdv$ potential. In fact, for $N>1$, there exist
nontrivial constraints on the set $\{z_j\}_{1\leq j\leq N}$ for
\eqref{3.2} to represent an algebro-geometric $\kdv$ potential. For
instance, if the $z_j$ in \eqref{3.2} are pairwise distinct, then
Airault, MacKean, and Moser \cite{AMM77} proved that
\begin{equation}
\sum_{\substack{j^\prime=1\\ j^\prime\neq j}}^N
\frac{1}{(z_{j}
-z_{j^\prime})^{3}}=0 \quad
   \text{for $j=1,\dots,N$} \lb{3.8aa}
\end{equation}
are necessary conditions for $q$ in \eqref{3.2} to be a stationary
KdV potential. In the
case of collisions {\rm (}i.e., if $s_{\ell_0} >1$ for some
$1\leq\ell_0\leq M${\rm )}  the necessary constraints on
$\{\zeta_\ell\}_{1\leq\ell\leq M}$ are more involved than in the
nondegenerate case above and a complete description
of all constraints were originally obtained by Duistermaat and Gr\"unbaum
\cite{DG86} in 1986. An alternative proof of their result will be given
in Theorem~\ref{t3.6} below. \\
{\rm(}ii{\rm)} In connection with
Theorem~\ref{t3.1}\,(ii) one might naively expect  that any decomposition
of $g(g+1)=\sum_{\ell=1}^M s_\ell(s_\ell+1)$ can actually be realized for
some choice of $\{\zeta_\ell\}_{1\leq \ell\leq M}$ with
$\zeta_\ell\neq\zeta_{\ell^\prime}$ for $\ell\neq\ell^\prime$.  However,
the simple counterexample
$q(z)=-6(z-\zeta_1)^{-2}-6(z-\zeta_2)^{-2}$, which satisfies
$\kdv_3(q)=-5670(\zeta_1-\zeta_2)^2(\zeta_1+\zeta_2-2z)(z-\zeta_1)^{-6}
(z-\zeta_2)^{-6}$, quickly destroys such hopes. \\
{\rm(}iii{\rm)} Strictly speaking, the version of Theorem~3.1\,(v)
proven in \cite{We99} assumes in addition to $q$ being rational, that
$q$ is bounded at infinity. However, assuming that
$$
q(z)\underset{z\to\infty}{=}\alpha z^k + O(z^{k-1}) \, \text{ for some
$\alpha\neq 0$ and $k\in\bbN$,}
$$
a simple inductive argument using \eqref{A.1} proves
$$
\hat f_j^\prime(z)=\frac{k\alpha^j}{2}\bigg(\prod_{\ell=1}^{j-1}
\frac{2\ell+1}{2\ell} \bigg)z^{jk-1} +O(z^{jk-2}), \quad j\geq 1,
$$
using the usual convention (for $j=1$) that products over empty sets
are put equal to one. Thus, since
$\hat f_j^\prime$ cannot vanish in this case, a rational
$q$ unbounded at infinity cannot satisfy any of the stationary KdV
equations (cf.~\eqref{A.6}). }
\end{rem}
Before we discuss
additional facts, we briefly pause and mention some of the ingredients
entering the proof of items (i)--(v) in Theorem~\ref{t3.1}. We start with
a fairly complete treatment of item (iii) and for simplicity of notation
put $q_\infty=\zeta_1=0$ and
\begin{equation}
q_g(z)=-g(g+1)z^{-2}, \quad g\in\bbN, \,\, z\in\bbC\backslash\{0\}.
\lb{3.8b}
\end{equation}
{}From \cite[Ch.~10]{AS72} one infers that
($E\in\bbC\backslash\{0\}$, $z\in\bbC$)
\begin{align}
\psi_\pm (E,z)=\Bigg(\sum_{k=0}^g \frac{(g+k)!}
{k!(g-k)!}(\pm 2E^{1/2}z)^{-k}\Bigg)e^{\mp E^{1/2}z},  \lb{3.9}
\end{align}
are linearly independent solutions of
$\psi''+(q_g-E)\psi=0$, $E\in\bbC\backslash\{0\}$.
Thus, one concludes that
\begin{equation}
\psi_+(E,z)\psi_-(E,z)=\prod_{j=1}^g\Big(1-\frac{\kappa_j}{Ez^2}\Big)
\text{ for some } \kappa_j\in\bbC, \,\, 1\leq j\leq g. \lb{3.10a}
\end{equation}
Hence a comparison with \eqref{A.10}--\eqref{A.13},
\eqref{A.16b}--\eqref{A.16g} yields
\begin{equation}
\hat F_g(E,z)=\prod_{j=1}^g \big(E-\mu_j(z)\big), \quad
\mu_j(z)=\kappa_jz^{-2}, \,\, 1\leq j\leq g, \lb{3.11}
\end{equation}
where $\hat F_g(E,z)$ denotes the polynomial of degree $g$
with respect to $E$ associated with $q_g(z)$ in \eqref{3.8b}, as
introduced in the appendix. Thus, $q_g(z)$ is a $\kdv$ potential
satisfying $\widehat \kdv_g (q_g)=0$
for a particular set of constants $\{c_\ell\}_{1\leq\ell\leq g}$ in
\eqref{A.8}. However, taking into account the  simple form of
$q_g(z)$ in \eqref{3.8b}, homogeneity considerations in connection
with the corresponding $\hat f_j$ and \eqref{A.17} then
yield in the special case $q(z)=q_g(z)$,
\begin{align}
& c_\ell=0, \quad 1\leq \ell\leq g, \lb{3.13} \\
& \hat F_g(E,z)=F_g(E,z), \quad \hat f_j(z)=f_j(z), \quad
1\leq j\leq g, \lb{3.14} \\
& f_j(z)=d_j z^{-2j} \text{ for some } d_j\in\bbC\backslash\{0\},
\,\, 1\leq j\leq g, \lb{3.15} \\
& f_{k+1}(z)=0, \quad \sKdV_k(q_g)=0, \quad k\geq g,
\lb{3.16} \\
& y^2=E^{2g+1}, \text{ that is, $\hat E_m=0$, \,\, $0\leq m\leq 2g$}
\lb{3.17}
\end{align}
(and of course $c_0=\hat f_0(z)=f_0(z)=1$). This yields item (iii)
and part of item (iv). Since $q$ in \eqref{3.2} and \eqref{3.5}
in the special case $q_\infty=0$ satisfies
$q(z)\underset{|z|\to\infty}{=} 2Nz^{-2}\big(1
+ O\big(|z|^{-1}\big)\big)$,
one infers that $f_{k+1}=0$ for some $k\in\bbN$ can only happen if
$N=k(k+1)/2$ for some $k\in\bbN$.
This illustrates $N=g(g+1)/2$ and \eqref{3.4b}. Item (v) in \cite{We99}
follows from a careful combination of Frobenius theory for second-order
linear ordinary differential equations in the complex domain, Halphen's
theorem, Theorem~\ref{t1.6} (for $n=2$), and some of the algebro-geometric
formalism briefly sketched in the appendix. As a by-product of a
proof of item (v) one shows that
$\psi''(z)-cz^{-2}\psi(z)=E\psi(z)$, $z\in\bbC\backslash\{0\}$
has a meromorphic fundamental system of solutions for all $E\in\bbC$ if
and only if $c\in\bbC$ is of the special form $c=s(s+1)$ for some
$s\in\bbN_0$.
This illustrates why collisions necessarily must happen as described in
\eqref{3.4a}. This fact was already known to Kruskal \cite{Kr74} in 1974.
That $q$ in \eqref{3.2}, \eqref{3.5}, and
\eqref{3.6} are all isospectral $\kdv$ potentials, that is, they all
belong to the same algebraic curve \eqref{3.8} (assuming \eqref{3.2} and
\eqref{3.5} satisfy the additional restrictions to make them
algebro-geometric $\kdv$ potentials, of course) can be shown by several
methods. Either by invoking time-dependent $\kdv$ flows as in
\cite{AMM77}, or by commutation techniques (i.e., Darboux-type
transformations) as in \cite{AM78}, \cite{EK82}, \cite{Mo77}, \cite{Oh88}
(cf.~also \cite{GH99b}). This fact also follows from the results in
\cite{We99}. Finally, identifying $\psi_\pm(E,z)/\psi_\pm(E,z_0)$ with
the two branches of the Baker-Akhiezer function $\psi(P,z,z_0)$,
$P=(E,y)$ in
\eqref{A.16c}, a combination of \eqref{A.10},
\eqref{A.16f}, and the normalizations
\begin{equation}
\lim_{|z|\to\infty} \psi_\pm(E,z)\exp(\mp E^{1/2}z)=1, \quad
\lim_{|E|\to\infty} \psi_\pm(E,z)\exp(\mp E^{1/2}z)=1, \no
\end{equation}
then proves
$\psi_+(E,z)\psi_-(E,z)=E^{-g}F_g(E,z)=\prod_{j=1}^g \bigg(1-
\frac{\mu_j(z)}{E}\bigg)$, and hence \eqref{3.8a}.

Finally, we study the precise restrictions on the set of
poles $\{z_j\}_{1\leq j\leq N}=\{\zeta_\ell\}_{1\leq\ell\leq M}$ for $q$
in \eqref{3.5} to be a $\kdv$ potential.
\begin{lemm} \label{l3.4}
Suppose the function $q$ has a Laurent expansion about the point
$z_0\in\bbC$ of the type
\begin{equation}
q(z)=\sum_{j=0}^\infty q_j (z-z_0)^{j-2}, \lb{3.30}
\end{equation}
where $q_0=-s(s+1)$ and, without loss of generality, $\Re(2s+1)\geq0$.
Define for $\sigma\in\bbC$,
\begin{align}
f_0(\sigma)&=-\sigma(\sigma-1)-q_0=(s+\sigma)(s+1-\sigma), \lb{3.31}\\
c_0(\sigma)&=\prod_{j=1}^{2s+1} f_0(\sigma+j), \,\,
   c_j(\sigma)=\frac{\sum_{m=0}^{j-1}
     q_{j-m} c_m(\sigma)}{f_0(\sigma+j)}, \;\, j\in\bbN, \label{crec} \\
   w(\sigma,z)&=\sum_{j=0}^\infty c_j(\sigma) (z-z_0)^{\sigma+j},
\lb{3.33} \\
   v(\sigma,z)&=\frac{\partial w}{\partial\sigma}(\sigma,z)
   =\sum_{j=0}^\infty \left(\frac{\partial c_j}{\partial\sigma}
    +c_j\log(z-z_0)\right) (z-z_0)^{\sigma+j}. \lb{3.34}
\end{align}
If $2s+1$ is not an integer, then $y''+qy=0$
has the linearly independent solutions $y_1=w(s+1,\cdot)$ and
$y_2=w(-s,\cdot)$. If $2s+1$ is an integer, then $y''+qy=0$ has the
linearly independent solutions $y_1=w(s+1,\cdot)$ and $y_2=v(-s,\cdot)$.

Moreover, $y''+qy=0$ has a meromorphic fundamental system of solutions
near $z_0$ if and only if $s\in\bb N_0$ and $c_{2s+1}(-s)=0$.
\end{lemm}
This is a classical result in ordinary differential equations (cf., e.g.,
\cite{In56}, Chs.~XV, XVI). A recent proof can be found in
Section 3 of
\cite{We99}.
\begin{defi} \lb{d3.5}
Let $q$ be a rational function. Then $q$ is called a \textit{Halphen
potential} if it is bounded near infinity and if $y''+qy=Ey$ has a
meromorphic fundamental system of solutions {\rm(}w.r.t.~$z${\rm)} for
each value of the complex  spectral parameter $E\in\bbC$.
\end{defi}
Of course every constant is a Halphen potential. Moreover, by
Theorem~\ref{t3.1}\,(v), $q$ is a Halphen potential if and only if it
is a rational KdV potential (i.e., if and only if it satisfies one and
hence infinitely many of the equations of the stationary KdV hierarchy).
\begin{theorem} \lb{t3.6}
Let $q$ be a nonconstant rational function. Then $q$ is a Halphen
potential if and only if there are $M\in\bbN$, $s_\ell\in\bbN$,
$1\leq\ell\leq M$, $q_\infty\in\bbC$, and pairwise distinct
$\zeta_\ell\in\bbC$, $\ell=1,\dots,M$, such that
\begin{equation}
q(z)=q_\infty-\sum_{\ell=1}^M s_\ell(s_\ell+1)(z-\zeta_\ell)^{-2}
\lb{3.35}
\end{equation}
and
\begin{equation}
\sum_{\substack{\ell^\prime=1\\ \ell^\prime\neq \ell}}^M
\frac{s_{\ell^\prime}(s_{\ell^\prime}+1)}{(\zeta_{\ell}
-\zeta_{\ell^\prime})^{2k+1}}=0 \quad
   \text{for $k=1, ..., s_{\ell}$ and $\ell=1,\dots,M$.} \lb{3.36}
\end{equation}
Moreover, $q$ is a rational KdV potential if and only if $q$ is of
the type \eqref{3.35} and the constraints \eqref{3.36} hold. In
particular, for fixed $g$, the constraints \eqref{3.36} characterize the
isospectral class of all rational KdV potentials associated with the
curve $y^2=(E-q_\infty)^{2g+1}$, where $g(g+1)=\sum_{\ell=1}^M
s_\ell(s_\ell+1)$.
\end{theorem}

\begin{proof}
By Theorem~\ref{t3.1}\,(v), it suffices to prove the characterization of
Halphen potentials. Suppose that
$q$ is a nonconstant Halphen potential. Then a pole
$z_0$ of $q$ is a regular singular point of $y''+qy=Ey$ and hence
$$
q(z)-E=\sum_{j=0}^\infty Q_j (z-z_0)^{j-2}
$$
in a sufficiently small neighborhood of $z_0$,
where $Q_2$ is a first order polynomial in $E$, while $Q_j$ for $j\neq2$
are independent of $E$. The indices associated with $z_0$, defined
as the roots of $\sigma(\sigma-1)+Q_0=0$ (hence they are
$E$-independent), must be distinct integers whose sum
equals one. We denote them by $-s$ and $s+1$ where $s>0$ and note that
$Q_0=-s(s+1)$. We intend to prove that $Q_{2j+1}=0$ whenever
$j\in\{0,...,s\}$ by applying Lemma \ref{l3.4}. Proceeding by way of
contradiction, we thus  assume that for some nonnegative integer
$k\in\{0,...,s\}$,
$Q_{2k+1}\neq 0$ and $k$ is the smallest such integer.

We note that $f_0(\cdot+j)$ are positive in $(-s-1,-s+1)$ for
$j=1,...,2s$, whereas $f_0(\cdot+2s+1)$ has a simple zero at
$-s$ and its derivative is negative at $-s$. Next one defines
\begin{equation}
\gamma_0(\sigma)=\prod_{j=1}^{2s+1} f_0(\sigma+j) \quad\text{and}\quad
   \gamma_1(\sigma)=\prod_{j=2}^{2s+1} f_0(\sigma+j). \lb{3.38}
\end{equation}
$\gamma_0$ and $\gamma_1$ have simple zeros at $-s$ and
and $\gamma_0'(-s)$ and $\gamma_1'(-s)$ are negative.

The functions $c_0=\gamma_0$ and $c_1=Q_1\gamma_1$ are polynomials
with respect to $E$. Actually, $c_0$ has degree zero in $E$ and $c_1$ is
constant but might equal zero. Hence the relations \eqref{ce},
\eqref{co}, and
\eqref{gamma} below are satisfied for $j=1$. Next we assume that for some
integer $\ell\in\{1, ..., s\}$, the functions
$c_0$, ..., $c_{2\ell-1}$ are polynomials in $E$ and that the relations
\begin{align}
   c_{2j-2}(\sigma)&=\gamma_{2j-2}(\sigma) Q_2^{j-1}+ O(E^{j-2}),
    \label{ce}\\
   c_{2j-1}(\sigma)&=\begin{cases}
    \gamma_{2j-1}(\sigma) Q_{2k+1} Q_2^{j-k-1}+ O(E^{j-k-2}),
     &j-1\geq k,\\
    0,&j-1<k, \end{cases} \label{co} \\
    \gamma_{2j-2}(-s)&=\gamma_{2j-1}(-s)=0, \quad
   \gamma_{2j-2}'(-s),\gamma_{2j-1}'(-s)<0 \label{gamma}
\end{align}
are satisfied for $1\leq j\leq\ell$. Using the recursion relation
\eqref{crec} we then obtain that $c_{2\ell}(\sigma)$ and
$c_{2\ell+1}(\sigma)$ are polynomials in $E$ and that
\begin{align*}
   c_{2\ell}(\sigma)
    &=\frac{\gamma_{2\ell-2}(\sigma)}{f_0(\sigma+2\ell)} Q_2^\ell
    +O(E^{\ell-1}),\\
   c_{2\ell+1}(\sigma)
    &=\begin{cases}
     \frac{\gamma_{2\ell-1}(\sigma)+\gamma_{2(\ell-k)}(\sigma)}
      {f_0(\sigma+2\ell+1)}
       Q_{2k+1} Q_2^{\ell-k}+O(E^{\ell-k-1}),&\ell\geq k, \\
     0,&\ell<k. \end{cases}
\end{align*}
Letting $\gamma_{2\ell}=\gamma_{2\ell-2}/f_0(\cdot+2\ell)$ and
$\gamma_{2\ell+1}
=(\gamma_{2\ell-1}+\gamma_{2(\ell-k)})/f_0(\cdot+2\ell+1)$ we find that
the relations \eqref{ce}, \eqref{co}, and \eqref{gamma} are satisfied
for $j=\ell+1$. Hence an inductive argument proves that $c_{2s+1}$ is a
polynomial in $E$ and that
\begin{align*}
c_{2s+1}(\sigma)
    &=\frac{\gamma_{2s-1}(\sigma)+\gamma_{2(s-k)}(\sigma)}
    {f_0(\sigma+2s+1)}Q_{2k+1} Q_2^{s-k}+O(E^{s-k-1})  \\
&=\gamma_{2s+1}(\sigma)Q_{2k+1} Q_2^{s-k}+O(E^{s-k-1}).
\end{align*}
But both $\gamma_{2s-1}+\gamma_{2(s-k)}$ and $f_0(\cdot+2s+1)$ have
simple zeros at $-s$ so that $\gamma_{2s+1}(-s)$ is different from zero.
Lemma~\ref{l3.4} then shows that $y''+qy=Ey$ has a solution which is not
meromorphic whenever $E$ is not a root of the polynomial $c_{2s+1}(-s)$.
This contradiction proves our assumption $Q_{2k+1}\neq0$ wrong.

Since $Q_1=0$, we proved that if $q$ is a Halphen potential
with pairwise distinct poles $\zeta_1,\dots,\zeta_M$, then the principal
part of $q$ about any pole $\zeta_\ell$ is of the form
$-s_\ell(s_\ell+1)/(z-\zeta_\ell)^2$ for an appropriate positive integer
$s_\ell$. Since $q$ is bounded at infinity a partial fraction expansion
then proves \eqref{3.35}. This immediately implies that for
$z_0=\zeta_\ell$,
\begin{equation}
Q_{2k+1}=2k\sum_{\substack{\ell^\prime=1\\ \ell^\prime\neq \ell}}^M
\frac{s_{\ell^\prime}(s_{\ell^\prime}+1)}{(\zeta_\ell
-\zeta_{\ell^\prime})^{2k+1}}. \lb{3.40}
\end{equation}
This proves necessity of the conditions \eqref{3.35} and \eqref{3.36} for
$q$ to be a Halphen potential. To prove their sufficiency we now assume
that \eqref{3.35} and \eqref{3.36} hold. Then, if $z_0$ denotes any of
the points
$\zeta_\ell$, one infers that the corresponding
$c_{2s_\ell+1}(-s_\ell)=0$.  Lemma~\ref{l3.4} then guarantees that all
solutions of $y''+qy=Ey$ are meromorphic and hence that $q$ is a Halphen
potential.
\end{proof}
\begin{rem} \lb{r3.7}
{\em {\rm(}i{\rm)} We emphasize again that the necessary and sufficient
conditions on $\zeta_\ell$ for $q$ in \eqref{3.35} to be a rational KdV
potential were first obtained by Duistermaat and Gr\"unbaum \cite{DG86}
in their analysis of bispectral pairs of differential operators. Our
approach based on Halphen's theorem and a direct Frobenius-type analysis
is a bit more streamlined since we aim directly at rational KdV
solutions (and do not cover the case of the Airy equation) but there
are undoubtedly some similarities in both approaches.  \\
{\rm(}ii{\rm)} We note that the restrictions
\eqref{3.36} simplify in the absence of collisions, where $s_\ell=1$,
$1\leq\ell\leq N$. In this case \eqref{3.36} reduces to
$\sum_{j^\prime=1, j^\prime\neq j}^N (z_j -z_{j^\prime})^{-3}=0$, $1\leq
j\leq N$, which  represents the well-known locus introduced by Airault,
McKean, and Moser \cite{AMM77}. This locus generated considerable
interest, and especially its generalizations to elliptic $\kdv$
potentials and (elliptic) $\KP$ potentials, were intensively studied
{\rm(}cf., e.g.,
\cite{Ai78}, \cite{Ca78}, \cite{CC77}, \cite{DS99}, \cite{EE95},
\cite{EK94}, \cite{Ka95}, \cite{Kr78}--\cite{Kr74}, \cite{Pe94},
\cite{Sh94}, \cite{Sm94}, \cite{Wi98}{\rm)}. The
current derivation  of \eqref{3.36} properly extends this locus to the
case of collisions {\rm(}i.e., to cases where some of the
$s_\ell>1${\rm)}. Moreover, this appears to be the first systematic
derivation of this locus {\rm(}with or without collisions{\rm)} within a
purely stationary approach {\rm(}i.e., without involving special
time-dependent
$\kdv$ flows, etc.{\rm)}.\\
{\rm(}iii{\rm)} For $k=1$, conditions \eqref{3.36} coincide with the
necessary conditions at collision points found by Airault, McKean, and
Moser \cite{AMM77} in their Remark~1 on p.~113. However, since there
are additional necessary conditions in \eqref{3.36}
corresponding to
$k\geq 2$, this disproves the conjecture made at the end of the proof of
their Remark~1. \\
{\rm(}iv{\rm)} The genus
$g=2$ ($N=3$) example,
$\tilde q_2(z,t)=-6z(z^3+6t)(z^3-3t)^{-2}$, $t\in\bbC$, with
$z_j=(3t)^{1/3}\omega_j$,
$\omega_j=\exp(2\pi ij/3)$, $1\leq j\leq 3$, explicitly illustrates the
locus in \eqref{3.36}. One verifies that $\tilde q_2(t)$
satisfies the $k$th stationary $\kdv$ equation, $\sKdV_k(\tilde q_2(t))=0$
for all $k\geq 2$ and all
$t\in\bbR$, as well as the 1st time-dependent $\kdv$ equation
$\tilde q_{2,t}=4^{-1}\tilde q_{2,xxx}+2^{-1}3\tilde q_2\tilde q_{2,x}$
{\rm(}see, e.g., \cite{Ai78}, \cite{DG86}{\rm)}. }
\end{rem}
Extensions of the stationary formalism described in this section to
elliptic $\kdv$ potentials are in preparation.
\appendix{The stationary KdV hierarchy} \lb{sA}
\renewcommand{\theequation}{A.\arabic{equation}}
\setcounter{theorem}{0}
\setcounter{equation}{0}

In this section we review basic facts on the stationary KdV
hierarchy. Since this material is well-known, we confine
ourselves to a brief account. Assuming $q$ to be meromorphic
in $\bbC$, consider the recursion relation
\begin{equation}\lb{A.1}
\hat{f}_0(z)=1, \quad \hat{f}'_{j+1}(z)=4^{-1}
\hat{f}'''_j(z) +
   q(z)\hat{f}'_j(z)+2^{-1}q'(z)\hat{f}_j(z)
\end{equation}
for $j\in\bbN_0$ (with $\prime$ denoting differentiation with respect to
$z$ and
$\bbN_0=\bbN\cup\{0\}$) and the
associated differential expressions (Lax pair)
\begin{align}
L_2 &=\dfrac{d^2}{dz^2} + q(z), \lb{A.2} \\
\hat{P}_{2g+1}&=\sum^{g}_{j=0} \left[-\dfrac{1}{2}
\hat{f}'_j(z)+
\hat{f}_j(z) \dfrac{d}{dz}\right] L_2^{g-j},\quad g\in\bb N_0. \lb{A.3}
\end{align}
One can show that
\begin{equation}\lb{A.4}
\left[ \hat{P}_{2g+1}, L_2 \right]=2\hat{f}'_{g+1}
\end{equation}
([$\cdot,\cdot$] the commutator symbol) and explicitly computes
from \eqref{A.1},
\begin{equation}\lb{A.5}
\hat{f}_0=1, \; \hat{f}_1=2^{-1} q+c_1,\;
\hat{f}_2=8^{-1}
q'' + 8^{-1} 3q^2 +c_12^{-1} q+c_2, \text{ etc.},
\end{equation}
where $c_j\in\bbC$ are integration constants. Using the convention
that the corresponding homogeneous quantities obtained by setting
$c_{\ell}=0$ for $\ell=1,2,\ldots$ are denoted by $f_j$, that is,
\begin{equation}
f_j=\hat{f}_j\big|_{c_{\ell}=0, \, 1\leq\ell\leq j}\, , \quad j\in\bbN,
\lb{A.5a}
\end{equation}
one obtains
\begin{equation}
\hat f_j=\sum_{\ell=0}^j c_\ell f_{j-\ell}, \quad 0\leq j\leq g.
\lb{A.5b}
\end{equation}
The (homogeneous) stationary KdV
hierarchy is then defined as the sequence of equations
\begin{equation}\lb{A.6}
\sKdV_g(q)=2f'_{g+1} =0, \quad g\in\bb N_0.
\end{equation}
Explicitly, this yields
\begin{equation}\lb{A.7}
\sKdV_0(q)=q^\prime =0, \quad \sKdV_1(q) =4^{-1}q'''+
2^{-1} 3qq'=0, \text{ etc. }
\end{equation}
The corresponding nonhomogeneous version of $\sKdV_g(q)=0$ is
then defined by
\begin{equation}\lb{A.8}
\widehat \sKdV_g (q)=2\hat{f}'_{g+1}=2\sum^{g}_{j=0}c_{g-j}f'_{j+1}=0,
\end{equation}
where $c_0=1$ and $c_1,...,c_g$ are arbitrary complex constants.

If one assigns to $q^{(\ell)}=d^{\ell}q/dz^{\ell}$ the degree
$\deg(q^{(\ell)}) = \ell+2, \; \ell\in\bb N_0$, then the
homogeneous
differential polynomial $f_j$ with respect to $q$ turns
out to have
degree $2j$, that is,
\begin{equation}\lb{A.9}
\deg (f_j)=2j, \quad j\in\bb N_0.
\end{equation}
Next, introduce the polynomial $\hat{F}_g(E,z)$ in $E\in\bb C$,
\begin{equation}\lb{A.10}
\hat{F}_g(E,z) = \sum^{g}_{j=0} \hat{f}_{g-j}(z) E^j
=\prod_{j=1}^\g (E-\mu_j(z)).
\end{equation}
Since $\hat f_0(z)=1$,
\begin{align}
&-2^{-1} \hat F_g''(E,z)\hat F_g(E,z)+4^{-1}\hat
F_g'(E,z)^2+(E-q(z))\hat F_g(E,z)^2 \no \\
&=\hat{R}_{2g+1}(E,z) \lb{A.12}
\end{align}
is a monic polynomial in $E$ of degree $2g+1$.
However, equations \eqref{A.1} and \eqref{A.8} imply that
\begin{equation}\lb{A.11}
2^{-1} \hat{F}'''_g-2(E-q)\hat{F}'_g+q'\hat{F}_g=0
\end{equation}
and this shows that $\hat{R}_{2g+1}(E,z)$ is in fact independent of $z$.
Hence it can be written as
\begin{equation}\lb{A.13}
\hat{R}_{2g+1}(E)=\prod^{2g}_{m=0} (E-\hat{E}_m), \quad
\{\hat{E}_m\}_{0\leq m\leq 2g} \subset \bb C.
\end{equation}

By \eqref{A.4} the nonhomogeneous KdV equation \eqref{A.8} is
equivalent to the commutativity of $L_2$ and $\hat{P}_{2g+1}$.
This shows
that
\begin{equation}\lb{A.14}
[\hat{P}_{2g+1}, L_2 ]=0,
\end{equation}
and therefore, if $L_2 \psi=E\psi$, this implies that
$\hat P^2_{2g+1}\psi
=\hat R_{2g+1}(E)\psi$. Thus $[\hat{P}_{2g+1}, L_2 ]=0$
implies $\hat P^2_{2g+1} = \hat R_{2g+1}(L_2)$
by the Burchnall and Chaundy theorem. This illustrates the intimate
connection between
the stationary KdV equation $\hat{f}'_{g+1}=0$ in
\eqref{A.8} and the
compact (possibly singular) hyperelliptic curve $\hat\calK_g$ of
(arithmetic)
genus $g$ obtained upon one-point compactification of
the curve
\begin{equation}\lb{A.16}
\hat\calK_g\colon y^2=\hat{R}_{2g+1}(E)=\prod^{2g}_{m=0}(E-\hat{E}_m)
\end{equation}
by joining the point at infinity, denoted by $P_\infty$. Points
$P\in\hat\calK_\g\backslash\{P_\infty\}$ will be denoted  by $P=(E,y)$,
moreover, the involution (hyperelliptic sheet exchange map) $*$ on
$\hat\calK_\g$ is defined by
\begin{equation}
*\colon\hat\calK_{g}\to\hat\calK_{g}, \quad P=(E,y)\mapsto P^{*}=(E,-y),
\,  P_{\pm\infty}^{*}=P_{\mp\infty}.\lb{A.16a}
\end{equation}

Introducing the meromorphic function $\phi(\cdot,z)$ on $\hat\calK_g$,
\begin{equation}
\phi(P,z)=\big[y(P)+(1/2)\hat F^\prime_g(E,z)\big]/\hat F_g(E,z), \quad
P=(E,y)\in\hat\calK_g \lb{A.16b}
\end{equation}
and the stationary Baker-Akhiezer function $\psi(\cdot,z,z_0)$ by
\begin{equation}
\psi(P,z,z_0)=\exp\bigg(\int_{z_0}^z dz^\prime\,\phi(P,z^\prime)
\bigg), \quad P\in\hat\calK_\g\backslash\{P_\infty\}, \lb{A.16c}
\end{equation}
one infers (for $P=(E,y)\in\hat\calK_\g\backslash\{P_\infty\}$,
$(z,z_0)\in\bbC^2$)
\begin{align}
L_2\psi(P,\cdot,z_0)&=E\psi(P,\cdot,z_0), \lb{A.16d} \\
P_{2g+1}\psi(P,\cdot,z_0)&=y\psi(P,\cdot,z_0), \lb{A.16e} \\
\psi(P,z,z_0)\psi(P^*,z,z_0)&=\hat F_g(E,z)/\hat F_g(E,z_0),
\lb{A.16f}
\\ W(\psi(P,\cdot,z_0),\psi(P^*,\cdot,z_0))&=-2y(P)/\hat F_g(E,z_0),
\lb{A.16g}
\end{align}
where $W(f,g)(z)=f(z)g^\prime (z)-f^\prime (z)g(z)$ denotes the Wronskian
of $f$ and $g$. Thus, $\psi(P,z,z_0)$ and $\psi(P^*,z,z_0)$ are linearly
independent solutions of $L_2\psi=E\psi$ as long as $E\in\bbC
\backslash\{\hat E_m\}_{0\leq m\leq 2g}$. The two branches of
$\psi(P,z,z_0)$ will be denoted by $\psi_\pm(E,z,z_0)$, respectively.

The above formalism leads to the following standard definition.
\begin{defi} \label{dA.1}
Any solution $q$ of one of the stationary KdV
equations \eqref{A.8} is called an {\bf algebro-geometric KdV
potential}.
\end{defi}
For brevity of notation we will occasionally call such $q$ simply
$\kdv$ potentials.

Finally, denoting $\hat {\ul E}=(\hat E_0,\dots,\hat E_{2g})$,
consider
\begin{align*}
&\bigg(\prod_{m=0}^{2g} \bigg(1-\frac{\hat E_m}{z}\bigg)
\bigg)^{1/2}=\sum_{k=0}^{\infty}c_k(\hat {\ul E})z^{-k}, \\
&\text{where }\, c_0(\hat {\ul E})=1,\quad
c_1(\hat {\ul E})=-\frac12\sum_{m=0}^N \hat E_m, \,\, \text{ etc.}
\end{align*}
Assuming that $q$ satisfies the $g$th stationary
(nonhomogeneous) KdV equation \eqref{A.8}, the integration constants
$c_\ell$ in \eqref{A.5b} become a functional of the $\hat E_m$ in the
underlying curve \eqref{A.16} and one verifies
\begin{equation}
c_\ell=c_\ell(\hat {\ul E}), \quad \ell=0,\dots,g. \lb{A.17}
\end{equation}

{\bf Acknowledgment.}
We are indebted to Wolfgang Bulla for discussions on this subject.

\end{document}